\documentclass[twocolumn]{aastex631}
\usepackage{amsmath}
\usepackage[utf8]{inputenc}

\usepackage{breqn}
\begin{document}


\title{Conditions for accretion favoring an unmelted Callisto and a differentiated Ganymede}

\author[0000-0002-0786-7307]{Yannis Bennacer}
\affiliation{Aix- Marseille Universit\'e, CNRS, CNES, Institut Origines, LAM, Marseille, France \\ 38 Rue Frédéric Joliot Curie, 13013 Marseille, France}

\author[0000-0001-5323-6453]{Olivier Mousis}
\affiliation{Aix- Marseille Universit\'e, CNRS, CNES, Institut Origines, LAM, Marseille, France \\ 38 Rue Frédéric Joliot Curie, 13013 Marseille, France}
\affiliation{Institut Universitaire de France (IUF), France}

\author[0000-0002-4094-5581]{Marc Monnereau}
\affiliation{IRAP, University of Toulouse, CNRS, Toulouse, France}

\author[0000-0001-9275-0156]{Vincent Hue}
\affiliation{Aix- Marseille Universit\'e, CNRS, CNES, Institut Origines, LAM, Marseille, France \\ 38 Rue Frédéric Joliot Curie, 13013 Marseille, France}

\author[0000-0002-3289-2432]{Antoine Schneeberger}
\affiliation{Aix- Marseille Universit\'e, CNRS, CNES, Institut Origines, LAM, Marseille, France \\ 38 Rue Frédéric Joliot Curie, 13013 Marseille, France}



\begin{abstract}
Analysis of Callisto's moments of inertia, derived from Galileo's gravity data, suggests that its structure is not fully differentiated. This possibly undifferentiated state contrasts sharply with the globally molten state inferred in its counterpart, Ganymede, and poses unique challenges to theories of the formation and evolution of the Galilean moons. During their formation, both moons experienced multiple heating mechanisms, including tidal heating, radiogenic heating from short-lived radionuclides, accretional heating from impacts, and heat from the surrounding circumplanetary disk. Our study investigates the optimal conditions required to account for Callisto's partially differentiated state in contrast to Ganymede’s complete differentiation. We investigate crucial accretion parameters, such as the timing of accretion onset, the duration of accretion, and the impactor size distribution. We find that the observed dichotomy between Ganymede and Callisto can be attributed to similar formation conditions, assuming an identical impactor size distribution and composition in the Jovian circumplanetary disk. The key differences in the formation of Ganymede and Callisto are the disk temperature at their respective formation locations and their final radii. Our results indicate that both moons accreted gradually over more than 2 Myr, concluding at least 5.5 Myr after the formation of calcium-aluminum-rich inclusions in the protosolar nebula. Our model demonstrates that Callisto can remain undifferentiated despite accreting a substantial influx of kilometer-sized impactors—potentially contributing up to 30\% of the total mass inflow—while still allowing for the complete differentiation of Ganymede. 

\end{abstract}



\section{\textbf{Introduction}} \label{sec:intro}


The remarkable diversity of giant planet moons in our Solar System provides a unique opportunity to investigate the origin and evolution of ocean worlds, both within these systems and in the broader context of exoplanetary science. A particularly fascinating feature of the Jovian system is the disparity in the internal structures of its moons. Specifically, the outermost moon, Callisto, shows no sign of global melting, in stark contrast to the fully differentiated interiors of the inner moons--Io, Europa, and Ganymede.

Analysis of Callisto's moments of inertia, based on Galileo gravity data ($C/MR^2$ = 0.3549) \citep{ANDERSON2001157}, indicates that its structure is not fully differentiated. This incomplete separation of ice and rock suggests that Callisto has remained cold from its accretion phase to the present days. The undifferentiated state of Callisto, in contrast to the globally molten state observed in its counterpart Ganymede \citep{Kivelson1988,ANDERSON2001157}, imposes unique constraints on theories regarding the formation and evolution of the Galilean moons.

Radiogenic heating, especially from short-lived radionuclides such as $^{26}$Al, is thought to be a primary driver of ice--rock separation, possibly triggering thermal runaway early in the history of giant planet satellites \citep{LEWIS1971}. Accretional heating, the heat generated by impacts during accretion, has also been proposed as a mechanism for melting the ice within the ice--rock mixture \citep{SCHUBERT198146}. The contribution of accretional heating depends on factors such as the size and velocity of the impactors within the Jovian circumplanetary disk (CPD) and the growth timescale of the protosatellite. The undifferentiated nature of Callisto is often attributed to late formation, a long accretion timescale, and the predominance of small accreting particles (less than $\sim$100 m in size) \citep{2008Barr, BARR2010_Titan}. However, larger impactors may have been present in Callisto's formation region \citep{estrada2009formation,Batygin_2020,2008Barr}. The size distribution of impactors during accretion is uncertain and varies widely between formation models. Two main scenarios are considered for the formation of satellites: pebble accretion and oligarchic growth. In pebble accretion, gas drag favors the accumulation of centimeter-- to meter--sized particles, leading to satellite formation \citep{Ronnet_2017}. In contrast, the oligarchic growth scenario assumes kilometer-sized particles in the CPD that grow through collisions, eventually reaching the size of the Galilean satellites \citep{estrada2009formation,Batygin_2020}. Since oligarchic growth requires larger impactors, the collision energy is buried deeper in the moon. \citet{MONTEUX2014} investigated Callisto's accretional heating and found that a substantial population of impactors larger than one kilometer within the Jovian CPD would be incompatible with Callisto remaining undifferentiated. While pebble accretion seems plausible, there is evidence for the presence of large particles in the CPD: kilometer-sized satellites exist both within Io's orbit and beyond Callisto's orbit in the current Jovian system. These bodies may represent collisional debris from the past evolution of the CPD, or planetesimals trapped by the CPD that have undergone partial ablation \citep{Ronnet_Johansen}.

Given the presence of both small and large particles in the Jovian CPD, a key question arises: Could a substantial population of kilometer-sized impactors during accretion be consistent with a possibly undifferentiated state of Callisto, while still allowing for the full differentiation of Ganymede? This question is the focus of our study.

In this analysis, we present a thermal evolution model designed to identify the specific accretion conditions that could result in an unmelted Callisto while allowing for a fully differentiated Ganymede. Our model simulates the growth of a satellite embryo within the CPD, accounting for impacts from diverse particle populations (cisplanetary impactors) originating within the disk. Although high-velocity heliocentric particles are often cited as the primary contributors to the Galilean satellites' cratered surfaces \citep{ZAHNLE1998202}, the energy flux from cisplanetary impactors is more influential during the rapid accretion phase due to their higher collision frequency \citep{Squyres}. In addition to impact heating, our model includes radiogenic heating from short-lived isotopes, thermal energy from the surrounding Jovian CPD, and tidal dissipation. Key variables in our analysis include the onset time and duration of accretion as well as the size distribution of impactors.

Section \ref{sec:model} outlines the thermal evolution model and the transport equation, incorporating heat contributions from radionuclides, impacts, tidal dissipation, and interactions with the surrounding CPD. Section \ref{subsection:Melting&Diff} describes the criteria assumed for moon differentiation and provides estimates of the characteristic timescales for ice--rock separation and transport within their interiors. In Section \ref{sec:results}, we present the results of a nominal case, followed by an analysis of how varying parameters influence the outcomes. Section \ref{sec:discussion} is dedicated to interpreting our findings and discussing their implications for the primordial dichotomy between Callisto and Ganymede.

\section{\textbf{Model}} \label{sec:model}

In this section, we provide an overview of the thermal evolution model used in our analysis, highlighting the modifications and enhancements made to address the specific objectives of our research.

\subsection{Thermal evolution model}



We use the one-dimensional model developed by \cite{MONNEREAU2013} to simulate the thermal evolution of a growing moon, incorporating radiogenic heating as a key heat source. This heating is essential during the early history of satellites and can potentially drive global melting within their interiors. Specifically, short-lived radionuclides such as $^{26}$Al and $^{60}$Fe provide nearly all the radiogenic energy during accretion, while the contribution from long-lived radionuclides is negligible during this phase. We opt to neglect \(^{60}\)Fe, as its initial \(^{60}\)Fe/\(^{56}\)Fe ratio is two orders of magnitude lower than the initial \(^{26}\)Al/\(^{27}\)Al ratio \citep{MISHRA201490}, resulting in a significantly lower initial heating rate. The time-dependent heating rate \(Q(t)_{\text{rad}}\) of \(^{26}\)Al is given by:

\begin{equation}
Q(t)_{\text{rad}} = m_r \rho X_{\text{Al}}q_0 \text{exp}[-\lambda_{\text{26}}(t+t_{\text{start}})],
\end{equation}

\noindent where $\rho$ is the uniform density of the body, \(m_r\) is the rock mass fraction, \(\lambda_{\text{26}} = 9.68 \times 10^{-7} \, \text{yr}^{-1}\) is the decay constant of \(^{26}\)Al. The initial heating rate per kg of Al is \(q_0 = 2.06 \times 10^{-5} \, \text{W/kg}\) using an initial \(^{26}\)Al/\(^{27}\)Al ratio of $5.85 \times 10^{-5}$ \citep{Thrane2006}, which is relevant for carbonaceous chondrites. 

The heating rate value \(q_0\) is consistent with those derived in the studies by \citet{2008Barr} and \citet{Trinh_etal_2023}, although slight variations are found in the literature. For instance, \citet{CASTILLOROGEZ2007179} adopted a value of $q_0$ = 7.3 $\times$ 10$^{-6}$ W/kg. The mass fraction of Al in CI chondrites is denoted by $X_{\text{Al}}$ with CI abondances taken from \cite{Lodders_2003}. The newly accreted material layers produce heating associated with their formation time \(t + t_{\text{start}}\), where \(t_{\text{start}}\) is the time when satellite accretion begins, following a chosen timespan after the formation of calcium-aluminium inclusions (CAIs) in the protosolar nebula (PSN).

During accretion, the heat received by protosatellites is primarily transported by conduction within the growing moon. Protosatellites acquire heat primarily from their surface due to accretion energy, resulting in a temperature gradient that opposes the conditions required for convection. While radiogenic heating in the satellite's interior can drive core convection, this process is unlikely to occur on a timescale shorter than the satellite's accretion period \citep{SCHUBERT198146}. The onset time for convection in Ganymede and Callisto has been estimated to be on the order of $\sim 10^8$ years \citep{2008Barr}, based on the melting point viscosity of Ice V and Ice VI ($\eta \sim 10^{17} \text{ Pa s}$). Since its timescale, $\tau_{\text{conv}}$, is longer than the accretion timescale, $\tau_{\text{acc}} \sim 10^6$ yr \citep{SCHUBERT198146}, we do not expect subsolidus ice convection to significantly affect the accretion temperature profiles.


The thermal evolution of the growing satellite is described by a 1-D heat diffusion equation in spherical geometry with a time-dependent radius \citep{MONNEREAU2013}:
\begin{align}
    \rho c_p \left( \frac{\partial T}{\partial t} - \frac{r^{\prime} \dot{R}_s}{R_s} \frac{\partial T}{\partial r^{\prime}} \right) 
    &= \frac{1}{(r^{\prime} R_s)^2} \frac{\partial}{\partial r^{\prime}} 
    \left( k r^{\prime 2} \frac{\partial T}{\partial r^{\prime}} \right) + Q_{\text{rad}},
    \label{diffusion}
\end{align}

\noindent where $R_s$ is the radius of the growing moon, $\dot{R}_s$ is its time derivative (the growth rate), $r'$ is the non-dimensional radial coordinate. The mean specific heat is denoted by $c_p$, and $k$ represents the thermal conductivity described in Sec. \ref{growth}. The heat diffusion equation is numerically solved using the implicit Euler method on a grid with 10,000 radial meshes, where the spatial step size, $\Delta r$ increases over time to account for the satellite's growth. The time step in our simulations is adaptive, determined by the characteristic timescale of impact dynamics.

\subsection{Circumjovian disk}

In this study we investigate the interaction between accretion heating and the temporal evolution of the CPD temperature in order to understand how cooling of the disk during the later stages of accretion affects satellite formation. The CPD model used in this study is described in \cite{Schneeberger2025} and builds on the work of \cite{makalkin1995,makalkin2014} and \cite{heller2015}. This model assumes that the CPD actively accretes material from the PSN with a protosolar metallicity $Z/H = 2.45 \times 10^{-2}$ \citep{lodders2019} and forms shortly after the end of Jupiter's gas runaway accretion phase. Over time, the CPD undergoes significant transformations, evolving from a massive, hot disk to a late-stage, depleted and cold state. Early in its evolution and close to the planet, the CPD is optically thick and vertically adiabatic. However, at larger distances from the planet and in later evolutionary stages, the CPD becomes optically thin and vertically isothermal.

Since the model assumes that the CPD forms immediately after Jupiter's formation, it incorporates the planet's early high surface temperature, which can reach up to $\sim$2000~K due to accretion energy \citep{szulagyi2016}. As a result, Jupiter's radiation heats the surface of the disk when it is optically thick and the midplane when it becomes optically thin. The model also takes into account the geometry of the CPD, which can cast shadows on itself, creating cold regions. In addition to Jupiter's radiative heating, the model takes into account viscous stress heating, accretion heating, and the temperature of the surrounding PSN.

\begin{figure}[!ht]
\resizebox{9cm}{!}{\includegraphics[angle=0,width=1cm]{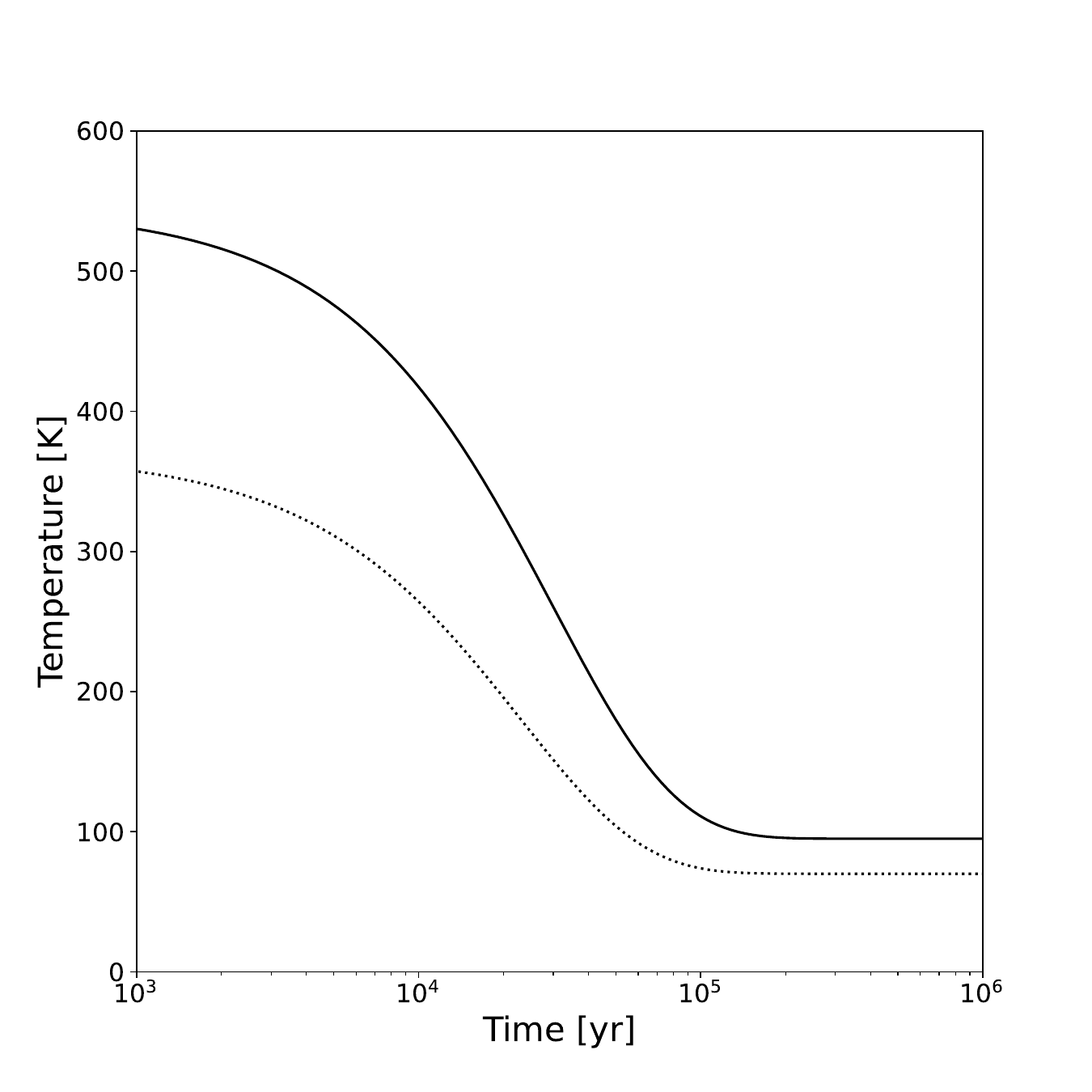}}
\caption{Evolution of the midplane effective temperature $T_e(t)$ in the Jovian CPD at the current orbits of Ganymede (solid line) and Callisto (dashed line). The CPD model assumes a metallicity $Z/H = 2.45 \times 10^{-2}$, a disk viscosity parameter of $10^{-3}$, a centrifugal radius of 50~$R_{\mathrm{J}}$, an ambient temperature $T_{\mathrm{neb}} = 40$~K, and a planetary temperature $T_{\mathrm{p}} = 2000$~K.
}
\label{figure:Disk_Temperature}
\end{figure}

 The surface temperature (or photosurface temperature) of the CPD is defined by:

\begin{equation}
    \begin{split}
    T_s^4(r) = & \frac{1 + \left( 2 \kappa_p(r,z_s) \Sigma_g(r) \right)^{-1}}{\sigma_{\mathrm{SB}}}\\
    &\times \left( F_{\mathrm{vis}} +  F_{\mathrm{acc}} + k_s F_{\mathrm{p}} \right) + T_{\mathrm{neb}}^4,
    \end{split}
\label{eq:surface_temperature}
\end{equation}

\noindent where $\Sigma_g(r)$ represents the gas surface density, $\sigma_{\mathrm{SB}}$ the Stefan-Boltzmann constant and $z_s$ the photospheric height. $T_{\mathrm{neb}}$, set to 40 K \citep{Schneeberger2025}, denotes the temperature of the surrounding PSN, and $\kappa_p$ is the radius-dependent opacity. $F_{\mathrm{vis}}$, $F_{\mathrm{acc}}$, and $F_{\mathrm{p}}$ are the heat fluxes produced by viscous stress, accretion of material onto the CPD, and Jupiter's radiative heating, respectively. The factor $k_s$ is the fraction of radiation absorbed by the disk, the rest being reflected and scattered away. The midplane effective temperature, $T_e$, is derived from the surface temperature, $T_s$, under the assumption that the inner region of the CPD is optically thick and adiabatic. Accordingly, $T_e$ is determined using a gray atmosphere radiative model \citep{makalkin2014,Schneeberger2025}. Figure \ref{figure:Disk_Temperature} shows the evolution of the CPD midplane temperature $T_e(t)$ at the present-day positions of Ganymede and Callisto, situated at 15 and 26.3 Jupiter radii, respectively. Accretion of the moons is assumed to begin when the snowline crosses their orbits, which are fixed in our study, typically occurring several tens of thousands of years after the CPD formation for Ganymede and Callisto. Since Ganymede would form in a more heated environment compared to Callisto, the condensation front migrates from Callisto's location to Ganymede's in only a few thousand years, a timespan significantly shorter than the timescale of the accretion phase, $\tau_{\text{acc}}$. The resolution method and detailed structure of the CPD are described in \cite{Schneeberger2025} and the references therein.

\subsection{Growing conditions}
\label{growth}

The accretion conditions of the Galilean moons are primarily determined by three factors: the lifetime of the CPD, the mass inflow rate, and the size of the bodies orbiting around the gas giant. The growth of the satellites occurs as they accrete particles from the CPD surrounding Jupiter, with a mass inflow rate depicted by the following relation:

\begin{equation}
\dot M_{\rm Sat} = \frac{M_{\rm Sat}}{\tau_{\rm acc}},
\end{equation}\label{eq:inflow_rate}

\noindent where $M_{\rm Sat}$ is the mass of the satellite and $\tau_{\rm acc}$ is the duration of accretion phase. As accretion timescales can differ widely between formation scenarios, we treat $\tau_{\rm acc}$ as a free parameter to explore a broad range of accretion conditions. This parameter ranges from 10 kyr, which aligns with the solids-enhanced minimum mass model \citep{MOSQUEIRA_a,MOSQUEIRA_b} to a few Myr, a value suitable for the longer accretion timescales derived from the classic gas-starved model \citep{2002CanupWard}.

We assume that Ganymede and Callisto accreted from the same source of pristine material within the Jovian CPD. The particles impacting the embryos have a radius $r_{\text{imp}}$ and the same mean density ($\rho \sim 1830$ kg/m$^{3}$) as Callisto, consisting of a uniform mixture of ice and rock, with a rock mass fraction $m_r=0.44$ \citep{MCKINNON1997540}. 

The thermal conductivity of the mixture is estimated as the mass-weighted harmonic average of the rock thermal conductivity, $k^{\text{rock}}= 4$ W/m/K \citep{Yomogida1983}, and the water ice thermal conductivity, $k^{\text{ice}} = 2.2$ W/m/K \citep{Waite2006}, with $k = (m_r/k^{\text{rock}} + (1-m_r)/k^{\text{ice}})^{-1} \sim$ 2.8 W/m/K. It is worth noting that the thermal conductivity of ice, $k^{\text{ice}}$, is inversely proportional to temperature, following the relation $k^{\text{ice}} \approx$ 600/T with T in K and $k^{\text{ice}}$ in W/m/K \citep{Klinger1980, Petrenko1999, McCord&Sotin2005}. However, this temperature dependence has a minimal impact in our study due to the relatively narrow temperature range considered (200--273 K) and the presence of rocks ($m_r$ = 0.44), which reduce the sensitivity of the effective thermal conductivity to temperature. The mean specific heat, $c_p \sim 1480 $ J/kg/K, is also inferred by the mass-weighted average, with $c_p^{\text{ice}} = 2100$ J/kg/K and $c_p^{\text{rock}} = 700$ J/kg/K \citep{2008Barr}.
 
The size distribution of impactors is given by the power law:

\begin{equation}
    \frac{dN}{dr_{\text{imp}}} \propto r_{\text{imp}}^{-\alpha}
    \label{eq:size_distribution}
\end{equation}

\noindent with the slope $\alpha$ ranging from 1 to 6 \citep{Squyres}. The radius of cisplanetary impactors, $r_{\text{imp}}$, can vary from a few centimeters to several hundred kilometers, depending on the disk model and the formation mechanism considered \citep{Ronnet_2017, estrada2009formation, Batygin_2020}. To account for the different sizes of bodies during accretion, we have considered three main particle populations, with each population contributing to the growth of the moon's radius $R_s(t)$ (see Table \ref{tab:Parameters} for the considered size range of impactors).

Small impactors are assumed to be large enough to deposit energy on the surface, but too small to penetrate significantly, leading to energy dissipation predominantly through radiation. Medium-sized impactors are large enough to deposit a fraction of their energy into the subsurface within a layer of thickness $\delta$, that exceeds the depth $\delta_{\text{cool}}$, subject to conductive cooling before accumulating a new layer \citep{2008Barr}. To enhance computational efficiency, continuous accretion is assumed for particles with radii below 1 kilometer. The accumulated mass from both medium and small particle populations is averaged and uniformly distributed across the surface using successive deposition layers.
In contrast, larger impactors are assumed to generate significant shock waves that can penetrate deeply into the icy-rocky structure of protosatellites \citep{MONTEUX2014}. These impactors are treated explicitly, with their radius and impact frequency randomly determined.

The mass fraction $x^i_{\rm m}$ of the \textit{i}$^{th}$ population is obtained by the ratio between the mass of \textit{i}$^{th}$ population $M^i$ and the total mass. The mass $M^i$ is derived by integrating the number of particles (Eq. \ref{eq:size_distribution}) multiplied by volume and density, yielding:

\begin{equation}
x^i_{\rm m} =  \frac{\int_{r^i_{\text{min}}}^{r^i_{\text{max}}} r_{\text{imp}}^{3-\alpha} \,dr_{\text{imp}}}{\int_{r_{\text{min}}}^{r_{\text{max}}} r_{\text{imp}}^{3-\alpha} \,dr_{\text{imp}}},\\
\text{ } \text{with} \text{ } x_{\rm m}^{si} + x_{\rm m}^{mi} + x_{\rm m}^{li} = 1,
\end{equation}

\noindent where $r^i_{\text{min}}$ and $r^i_{\text{max}}$ represent the size range of the \textit{i}$^{th}$ population (see Table \ref{tab:Parameters}), and superscripts $si$, $mi$, and $li$ denote the populations of small, medium-sized, and large impactors, respectively.


\begin{table}[ht]
\centering
\caption{Impactor populations}
\begin{tabular}{lcc}
\hline
\hline
Population & Size range & Energy deposit \\ \hline
Small    &    1 cm--100 m      &  Surface \\ 
Medium   &    100 m--1 km & Surface and subsurface \\ 
Large    &     1--100 km & Deep inside \\ 
\hline
\label{tab:Parameters}
\end{tabular}
\end{table}

Large impactors are treated individually, with impact probabilities determined using a Monte Carlo approach. For each incoming particle, the radius of the impactor is randomly generated based on the probability density function $ f(r_{\text{imp}}) = Cr^{-\alpha}_{\text{imp}}$ with C a constant. By reversing the cumulative distribution function $F_R(r)$, the radius of the next impactor is given by:
\begin{equation}
\begin{split}
&r_{\text{imp}} = \left(\frac{F_R(r)(1-\alpha)}{C} + r_{\text{min}}^{(1-\alpha)} \right) ^{\frac{1}{1-\alpha}} \\ 
&\text{with } C = \frac{1-\alpha}{r^{1-\alpha}_{\text{max}} -r^{1-\alpha}_{\text{min}} }
\end{split},
\end{equation}

\noindent where $F_R(r)$ is a random number drawn between 0 and 1. The latitude $\theta$ and longitude $\phi$ of each impact are also randomly generated to simulate random impact frequencies. We assume here an isotropic impact flux as inclinations and eccentricities of large particles are randomized by frequent collisions, preventing any preferred impact direction. The latitude and longitude are given by:

\begin{equation}
\begin{cases}
    \theta = \frac{180}{\pi}\arcsin(1 - 2r_1) \\
    \phi = 4 \times \frac{180}{\pi} \left| \arcsin(1 - 2 r_2) \right|
\end{cases},
\end{equation}

\noindent where $r_1$ and $r_2$ are two independent random numbers between 0 and 1. An impact occurs when the incoming particle crosses a reference impact location. Neighboring impacts are considered by taking a margin of one impact diameter, i.e. the outer edges of the particle must at least touch the impact site, which is defined by a set of latitude-longitude coordinates. This one-dimensional approach allows us to examine a random point on the protosatellite and study its internal structure.


When a large particle impacts the surface, a shock wave propagates from the impact site toward the center of the satellite. The initial kinetic energy of the impactor is converted into shock heating, internal energy produced by plastic work, and the kinetic energy of the ejecta \citep{Squyres}. We do not include the ejecta in the energy balance because the fraction of energy converted to kinetic energy is only between 7 and 10\% \citep{Okeefe1977}. We also neglect the post-shock impact heating caused by plastic deformation, as shock physics simulations have shown that the heat is primarily localized at the surface \citep{Kurosawa2021}.

When a shock wave propagates through a protosatellite, the peak shock pressure is concentrated within the isobaric core, a spherical region where the pressure is approximately uniform \citep{Croft1982}. The temperature increase $\Delta T$ within the isobaric core due to each impact is given by the following relation \citep{Monteux2007,MONTEUX2014}:

\begin{equation}
\Delta T = \frac{4\pi}{9} \frac{\gamma_{li}\rho G R_s^2}{h_m c_p},
\end{equation}   

\noindent where $R_s$ is the growing satellite radius and $G$ is the gravitational constant. The term $h_m = \frac{V_{\text{eff}}}{V_{ic}} \sim 5.8 $ \citep{MONTEUX2014} represents the volume effectively heated normalized by the volume of the isobaric core, with $V_{\rm ic}= 3V_{\rm imp}$ and $V_{\rm imp}$ being the volume of the impactor \citep{Senshu2002}. The parameter $\gamma_{li}$ denotes the fraction of the impactor's kinetic energy converted into shock heating, which ranges from 0.2 to 0.6 \citep{Okeefe1977} for low-velocity ($v_{\rm imp} \lesssim$ 5 km/s) and high-velocity ($v_{\rm imp} > 10$ km/s) impacts, respectively. Since this study only considers cisplanetary particles, we neglect the impactor velocity at infinity, resulting in $v_{\rm imp} \approx v_{\rm esc}$, where $v_{\rm esc}$ is the escape velocity. For Ganymede and Callisto, where $v_{\rm imp} < 3$ km/s, we adopt $\gamma_{li} = 0.2 $.

Outside the isobaric core, both the shock pressure and the thermal perturbations decay significantly via the following relation \citep{Monteux2007}:

\begin{equation}
    T(\bar{r}) = 
    \begin{cases}
     T + \Delta T & \quad \bar{r} \leq r_{\rm ic} \\ 
     T + \Delta T \left(\frac{r_{\rm ic}}{\bar{r}}\right)^m & \quad \bar{r} > r_{\rm ic}
    \end{cases},
\end{equation}

\noindent where $\bar{r}$ is the distance from the center of the isobaric core with radius $r_{\rm ic}$ and $m$ is an exponent ranging from 2.8 to 4. In this study, we adopt the median value $m = 3.4$ \citep{MONTEUX2014}. 

While large impactors are treated individually, small and medium impactors are uniformly added to the surface with an average mass inflow rate of $\dot m^i_{\rm imp}~=~x^i_{\rm m} \dot M_{\rm Sat}$. At each time step, a thin uniform layer is deposited on the surface with a thickness given by:

\begin{equation}
\delta^{i} = \left( \frac{3 \dot m^i_{imp}\Delta t}{\rho 4\pi} + R_s^3\right) ^{1/3} - R_s.
\end{equation}

Medium impactors heat the surface but also the subsurface, depositing a fraction of their energy uniformly to a depth of $\delta^{\rm mi}$. Assuming $v_{\rm imp}$ $\approx$ $v_{\rm esc}$, the temperature profile of the additional layer is given by the relation \citep{MONTEUX2014, SCHUBERT198146}:

\begin{equation}
T(R_s) = \frac{\gamma_{\rm mi}GM_s}{c_p R_s} + T_e,
\end{equation}

\noindent where $M_s$ is the mass of the growing moon, and $T_e$ is the time-dependent temperature of the disk around the gas giant. The fraction of kinetic energy converted into internal energy from medium impactors is $\gamma_{\rm mi} = 0.1$. The parameters $\gamma_{\rm mi}$ and $\gamma_{\rm li}$ differ from the parameter $h$ introduced by \cite{SCHUBERT198146}, which includes both the energy that goes into heating and the post-impact surface cooling (see discussion in \citealt{MONTEUX2014}).

The surface is continuously heated by small and medium impactors originating from the background disk around Jupiter. We assume these impactors are at the same temperature as the CPD gas and are small enough to be approximated by continuous accretion. Besides the gravitational energy deposited during the impact, the surface is also heated or cooled, depending on whether these bodies are hotter or colder than the surface. The time-dependent surface temperature $T_{\text{surf}}$, used as the outer boundary condition for the heat diffusion equation, is determined by the balance between the gravitational energy deposited by impactors, radiation to space, and heat conducted to the surface or toward the center of the growing body: 

\begin{equation}
\begin{split}
&\sigma (T_{\text{surf}}^4 - T_e^4)4\pi R_s^2 = k \frac{\partial T}{\partial r}4\pi R_s^2  \\
&\quad + \frac{GM_s \dot{M}_{\text{sat}}}{R_s} (x^{si}_{\rm m} + x^{mi}_{\rm m}(1 - \gamma_{mi})) \\
&\quad - c_p \dot{M}_{\text{sat}}(x^{mi}_{\rm m} + x^{si}_{\rm m})(T_{\text{surf}} - T_e).
\end{split}
\label{equation:Tsurf}
\end{equation}

\noindent Here, $\sigma$ is the Stefan-Boltzmann constant and $(1 - \gamma_{mi})$ represents the fraction of the initial kinetic energy of medium impactors that is not buried inside the satellite but is instead subject to radiative cooling. The left-hand side of the equation represents the difference between the radiation from the surface and the background disk. From left to right, the other terms of the equation describe heat conduction, gravitational potential energy delivered by small and medium impactors, and the heating (or cooling) induced by these particles. Larger impactors also heat the surface, but this energy is efficiently radiated away. After an elevation of temperature $\delta T_0$ caused by an impact, the surface temperature difference decreases over time according to $\delta T(t)/\delta T_0\approx \sqrt{\kappa}\rho c_p/4\sigma T_e^3\sqrt{\pi  t}$ \citep{Carslaw1959} for times longer than the cooling timescale  $t_{\text{cool}} = 10 \kappa(\frac{\rho c_p}{4\sigma T_e^3})^2$, where $\kappa \sim 10^{-6}$ m$^2$/s, $T_e = 200$ K, $\rho \sim 1830$ kg/m$^3$ and $c_p \sim 1480$ J/kg/K. After approximately $t_{\text{cool}} \sim 250$ days, the temperature difference drops to less than 20\% of the initial temperature increase $\delta T_0$, and after 10 years, our average time step $dt$, it falls below 5\%. Since our study does not focus on the detailed short-term evolution of surface temperature, the fraction of the initial kinetic energy of large impactors lost through radiative cooling can be neglected.

As protosatellites grow, tidal forces increasingly dissipate energy within them, contributing significantly to their heat budget. The tidal dissipation in a homogeneous body in synchronous rotation with \textit{small} eccentricity $e$ follows the prescription \citep{Wisdom2004}:

\begin{equation}
\frac{dE}{dt} = \frac{21}{2} \frac{k_2}{Q} \frac{G M^2_{\text{Jup}} n R^5_s e^2}{a^6}
\label{equation:tidal}
\end{equation}

\noindent where $k_2$ is the Love number, $Q$ is the tidal dissipation parameter, $n$ is the orbital mean motion, and $a$ is the semimajor axis. The factor $\frac{k2}{Q}$ represents the deformability of the satellite. For Ganymede and Callisto, we assume $\frac{k_2}{Q} \approx 0.05$ \citep{2020Downey, Malhotra1991} and adopt constant orbital parameters $(a, e)$ during accretion, assuming that the satellites formed in situ. Note that an additional expression for tidal dissipation in a body undergoing non-synchronous rotation has been provided by \cite{WISDOM2007}, which shows that the dissipation in the case of synchronous rotation is greater than or equal to that in the case of asymptotic non-synchronous rotation. The use of Eq. \ref{equation:tidal} is justified by the fact that radiogenic heating dominates by several orders of magnitude during the accretion phase, as shown below, and because we assume in situ formation of the moons. However, due to the significant uncertainties associated with tidal dissipation, spin states, and orbital dynamics during this period, it is important to emphasize that Equation \ref{equation:tidal} represents an end-member scenario of minimal tidal heating. In this study, we have adopted several simplifying assumptions, including synchronous rotation, the neglect of dissipation in fluid layers, and the use of a constant dissipation factor that does not account for variations in temperature, composition, or pressure.

\section{Melting and Differentiation}\label{subsection:Melting&Diff}

\begin{table*}[!bt]
\centering
\caption{Ice Phases and corresponding coefficients \citep{Leliwa2002}} 
\begin{tabular}{lcccc}
\hline
\hline
Ice Phases & Pressure Range [MPa] & $T_{0}$ [K] & $A$ [K/Pa] & $B$ [K/Pa$^{2}$] \\
\hline 
Ice I & 0 $<$ $P$ $\leq$ 209 & 273.2 & -7.95 $\times$ 10$^{-8}$ & -9.6 $\times$ 10$^{-17}$ \\ 
Ice III & 209 $<$ $P$ $<$ 344 & 247.7 & 2.38 $\times$ 10$^{-8}$ & 0 \\ 
Ice V & 344 $<$ $P$ $<$ 626 & 242.5 & 4.9 $\times$ 10$^{-8}$ & 0 \\ 
Ice VI & 626 $<$ $P$ $<$ 2150 & 190.3 & 1.54 $\times$ 10$^{-7}$ & -3.43 $\times$ 10$^{-17}$ \\ 
Ice VII & $P$ $>$ 2240 & 149.8 & 9.14 $\times$ 10$^{-8}$ & 0 \\   
\hline
\end{tabular}
\label{tab:Ice_phases}
\end{table*}

To determine if melting occurs, we compare the temperature profiles within the newly formed satellite to the pressure-dependent melting temperature of water ice. To do so, we account for the different phases of ice as the pressure $P(r,t)$ increases over time in a given layer of radius $r$:

\begin{equation}
P(r,t) = \frac{2 \pi}{3} G\rho^2(R^2_s(t)-r^2).
\end{equation}

\noindent We assume a uniform density and zero pressure at the surface. 

The time-dependent liquidus temperature for a solid mixture of ammonia and water, including high-pressure phases, is determined using the formulation provided by \cite{Leliwa2002}:

\begin{equation}
\begin{split}
&T_m(P(r,t),X_{\text{NH$_3$}}) \approx T_0 + AP \\
& + BP^2 -CX_{\text{NH$_3$}} -DX_{\text{NH$_3$}}^2,
\end{split}
\end{equation}

\noindent where $T_0$, $A$ and $B$ are the coefficients for the different phases of pure water ice given in Table \ref{tab:Ice_phases}, and taken from \cite{Leliwa2002}. $C$ and $D$ are coefficients with values of 53.8 and 650, respectively. Since other volatiles could condense in the CPD, ammonia and other salts could be present in the ice mixture, lowering the melting point of water and allowing a subsurface ocean to remain stable beneath a convecting ice shell \citep{Spohn_Schubert_2003}. However, the presence of ammonia raises the question of whether its inclusion is consistent with the formation of Callisto without significant melting. To investigate this effect, specifically the effect of NH$_3$ on an undifferentiated Callisto, we have included this species in the ice mixture, where $X_{\text{NH$_3$}}$ represents the mass fraction. Since the phase diagram of NH$_3$ at high pressure remains poorly constrained \citep{Mousis2002}, we consider ammonia only at the liquidus temperature in the outer layers of the protosatellite.

If melting occurs during accretion, rock particles can settle to the bottom of the resulting localized pool of liquid water, leading to local differentiation. Specifically, an impact that melts the ice produces a transient pool of liquid water in which the denser rock particles may separate and settle at the base, provided they can sink before the water refreezes. Modeling the rock particles as spheres of radius $r_{\text{rock}}$, their settling velocity can be estimated using Stokes' law. This yields a sinking timescale $t_{\text{sink}}= 9\eta_ld_{\text{pool}}/2g(\rho_r-\rho_l)r^2_{\text{rock}}$ \citep{2010Barr_Dichotomy} where $d_{\text{pool}}$ is the depth of the pool, $\eta_l=10^{-3}$ Pa.s is viscosity of liquid water, $\rho_r=3000$ kg/m$^3$ is the rock density, $\rho_l= 1000$ kg/m$^3$ is the liquid water density, and $g = 1.24$ m/s$^2$ is the gravity on Callisto. For an impactor of radius \( r_{\text{imp}} = 1 \) km, which melts a volume corresponding to \( V_{\text{ic}} = 3V_{\text{imp}} \) \citep{Senshu2002}, where $V_{\text{ic}}$ is the volume of the isobaric core, the pool depth is given by $d_{\text{pool}}=r_{\text{ic}}$, where $r_{\text{ic}}$ is the radius of the isobaric core. For small grains particles $r_{\text{rock}} = 1$ cm, the estimated timescale is $t_{\text{sink}}\sim 30$ s. This must be compared to the timescale for the pool to freeze $t_{\text{cryst}}\sim L\rho_lr_{\text{ic}}^2/k^{\text{ice}}\Delta T$ where $L$~=~$3 \times 10^5$ J/kg is the latent heat of water ice, $\Delta T$ is the temperature gradient between the cold surface and the depth of the pool. For $\Delta T = 150$ K, this yields $t_{\text{cryst}}\sim 2 \times10^4$ yr, consistent with published estimates of freezing times $(10^3$--$10^5$ yr) \citep{Michaut,Hesse&Castillo2019,Chivers_etal_2021,Naseem_2023}. Given that $t_{\text{sink}}\ll t_{\text{cryst}}$, even the smallest accreted and co-accreted particles would likely settle to the bottom of the melt pool well before it solidifies.

\begin{figure}[!ht]
\centering
\resizebox{9cm}{!}
{\includegraphics[angle=0,width=10cm]{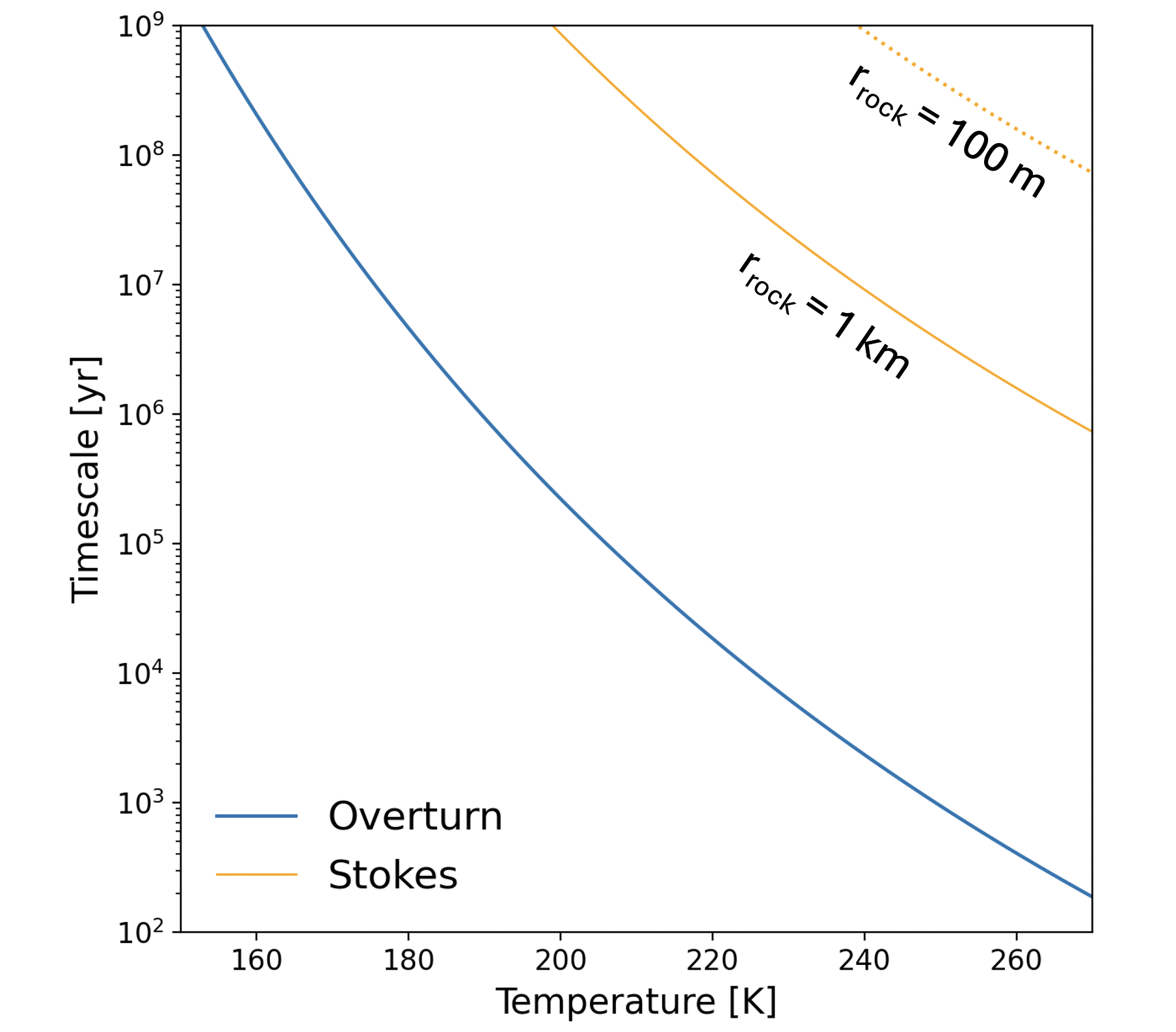}}
\caption{Stokes sinking and Overturn timescales versus typical mantle temperature (150--273 K). Both timescales are inferred using the viscosity of the ice rock mixture $\eta^{\text{mix}}=\eta^{\text{ice}}/f(\phi)$ and an Arrhenius-type relation for ice viscosity $\eta^{\text{ice}}=\eta^{\text{ref}}\text{exp}(A(T_{\text{ref}}/T-1)$. Timescales for Stokes sinking are displayed for different size of rock particles (100 m--1 km).}
\label{Timescale_Differentiation}
\end{figure}

Once separated from the ice and settled at the base of the melt pool, the layer of dense rock particles becomes gravitationally unstable and may undergo convective overturn driven by Rayleigh–Taylor (RT) instabilities \citep{RUBIN2014,Shoji&Kurita2014}. These instabilities arise when the dense, rock-rich layer overlies a less dense, undifferentiated region, or when an undifferentiated layer is subsequently deposited atop a refrozen, rock-depleted melt pool. In either case, the instability facilitates the continued descent of rock material toward the interior, contributing to partial differentiation of the moon. Following the studies of \citet{RUBIN2014} and \citet{Shoji&Kurita2014}, the timescale for the development of RT instabilities can be estimated by:
\begin{equation}
    \tau_{\text{RT}} =  \eta^{\text{mix}} \left( (n-1)^{(1/n)}\frac{C_{L\Delta \rho}}{2n}\Delta \rho g L_{\text{RT}} \right) ^{-1} \left( \frac{Z_0}{L_{\text{RT}}} \right) ^{(1-n)/n}
\end{equation}
where $\eta^{\text{mix}}$ is the viscosity of the ice-rock mantle, $C_{L\Delta \rho}\approx 0.76$ is a dimensionless constant that depends on the rheology and geometry of the system, and $n$ is the stress exponent, for which we adopt $n=1.8$, assuming that grain boundary sliding dominates the creep behavior of the ice \citep{RUBIN2014}. $Z_0$ represents the initial perturbation amplitude and $L_{\text{RT}}=nRT_0^2/E_a|dT/dz|$ the length-scale over which the viscosity varies significantly, with $T_0$ the temperature at the interface, $E_a$ an activation energy and $R=8.314$ J/mol/K the gaz constant. We assume $T_0=200$ K, $E_a = 49 \times 10^3$ J/mol, $|dT/dz|= 1.8$ K/km, and $Z_0=1$ km \citep{RUBIN2014}. The viscosity of the ice-rock mantle is modeled as $\eta^{\text{mix}}=\eta^{\text{ice}}/f(\phi)$ \citep{NAGEL2004402} where $\eta^{\text{ice}}$ is the ice viscosity which depends exponentially on temperature and $f(\phi)$ accounts for the influence of the rock fraction. For simplicity, we adopt an Arrhenius-type relation for the viscosity of ice, $\eta^{\text{ice}}=\eta^{\text{ref}}\text{exp}(A(T_{\text{ref}}/T-1)$, with $\eta^{\text{ref}}=10^{14}$ Pa.s, $T_{\text{ref}}=273$ K, and $A=$ 20 \citep{Shoji&Kurita2014}. The influence of the rock volume fraction is given by $f(\phi)=(1-\phi_0/\phi_{\text{CPL}})^\beta$, where $\phi_0=m_r\rho/\rho_r=0.27$ is the assumed volume fraction of rock, $\phi_{\text{CPL}}=0.74$ is the close--packing limit, and $\beta$ is an empirical exponent ranging from 1.55 to 2.5 \citep{NAGEL2004402}. For our study, we adopt the median value, $\beta$ = 2.0. Figure \ref{Timescale_Differentiation} illustrates the Rayleigh–Taylor overturn timescale $\tau_{\text{RT}}$ as a function of typical mantle temperatures (150–273 K). The figure clearly illustrates the strong temperature dependence of the instability timescale, which spans from approximately 1 Gyr to just a few hundred years across the range of mantle temperatures examined in this study. Our analyses suggest that gravitationally unstable, rock-rich layers are likely to overturn on timescales comparable to the duration of satellite accretion, particularly when mantle temperatures exceed $\sim$180 K, a threshold reached in all simulations.

Differentiation may also be facilitated by Stokes sinking, where ice rock separation occurs via solid--state creep of ice around the denser rock. A rock fragment could cross the satellite by Stokes flow over a timescale given by:

\begin{equation}
\tau_{\text{Stokes}} = \frac{9}{2}\frac{R_s\eta^{\text{mix}}}{(\rho_r-\rho_{\text{ice}})gr^2_{\text{rock}}},
\end{equation}

\noindent where $\rho_{\text{ice}} = 1400$ kg/m$^3$ represents the compressed density of various high-pressure ice phases \citep{2008Barr}. Figure \ref{Timescale_Differentiation} also presents the Stokes sinking timescale $\tau_{\text{Stokes}}$ as a function of typical mantle temperatures (150--273 K) and rock fragment sizes (100 m--1 km). This suggests that kilometer-scale particles (the typical size of the largest particles generated in our simulations) are unlikely to reach the base of the moon within accretion timescales unless the mantle approaches the melting temperature of water ice. Smaller particles (1 cm--100 m) are even less likely to settle before complete melting of the ice mantle occurs. However, it is important to note that Stokes sinking can also occur after rock has already separated from the ice within a melt pool. In this context, the size of a sinking fragment does not necessarily reflect that of the initially accreted particles. Once at the base of the melt pool, individual rock grains may accumulate and coalesce into larger fragments before the pool fully refreezes \citep{tonks2016impactinduceddifferentiationicybodies,BARR2010_Titan,2010Barr_Dichotomy}. This process could lead to the formation of significantly larger rock bodies than those initially accreted, thereby substantially reducing their Stokes sinking timescale. However, because this mechanism is highly sensitive to the melt volume and remains poorly constrained, it is not quantified and is excluded from the scope of this study.

Overall, our calculations indicate that efficient rock–ice separation likely begins once the liquidus is reached and melting initiates. Following decoupling from the ice, rock particles can migrate inward and contribute to core formation on timescales comparable to that of satellite accretion, primarily through convective overturn of gravitationally unstable layers. The gravitational energy released during this downward migration may significantly influence the satellite’s thermal evolution, as it is substantial—approximately $\sim$3$\times 10^5$ J/kg, nearly two orders of magnitude greater than the energy required to melt all the ice, estimated at $(1 - m_r)ML$ $\sim$5$\times 10^3$ J/kg. In this context, we assume that global melting and subsequent differentiation occur once the water liquidus is exceeded, an assumption also adopted by \citet{MONTEUX2014} and \citet{2008Barr}.
\section{\textbf{Results}} \label{sec:results}

\begin{figure*}[!ht]
\resizebox{\hsize}{!}{\includegraphics[angle=0,width=5cm]{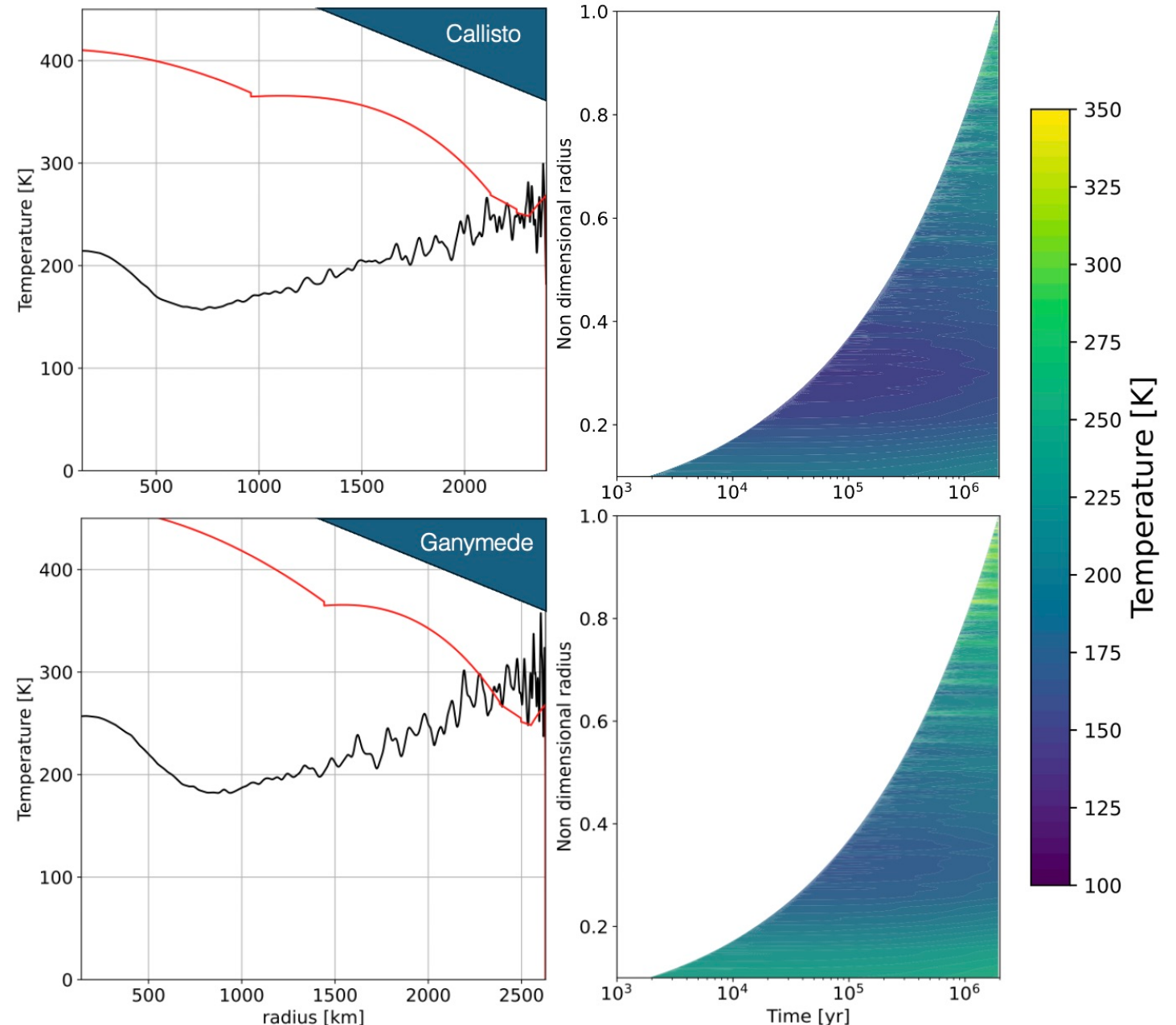}}
\caption{Simulation of the accretion of a protosatellite similar to Callisto and Ganymede, with parameters $\alpha=4$, $\tau_{\text{acc}}=2$ Myr, and $t_{\text{start}}=4$ Myr. The body begins accreting material at $t_{0}=t_{\text{start}}$, starting with an initial radius $R_{0}=100$ km. The simulation stops when the moon's radius reaches that of Callisto $R_f=2410$ km and Ganymede $R_f=2630$ km at $t_{\rm end}=t_{\text{start}}+ \tau_{\text{acc}}$.  {\it Left panel:} temperature profile radius (black line) within the protosatellite at the end of the accretion phase. The red line represents the melting curve $T_m(P,X_{\rm NH_3})$ for water ice with 5\% NH$_3$. The layer where the temperature profile exceeds the melting curve undergoes melting, where the percentage of material above the melting point of water ice (in volume) is 2.2 \% and 10.4\% respectively for Callisto and Ganymede. {\it Right panel:} time evolution of the temperature distribution during the accretion phase.}
\label{figure:Callisto_example}
\end{figure*}

Figure \ref{figure:Callisto_example} shows an example of temperature profile inside a satellite similar to Callisto and Ganymede, referred to as our nominal model, at the end of the accretion phase (left panel) and over time (right panel), including radiogenic and tidal heating. The nominal model is characterized by the critical accretion parameters: $t_{\text{start}}$, $\tau_{\text{acc}}$, and $\alpha$, respectively the onset time of accretion, the duration of the accretion phase, and the slope of the impactor size distribution. In our nominal model, we assume $\alpha = 4$, leading to a large impactor contribution of about $x^{li}_{\rm m}=30$ \% of total mass. The heat deposited deep within the growing moon by these large impactors is slow to dissipate, as evidenced by the temperature spikes shown in the left panel of Fig. \ref{figure:Callisto_example}. Since conduction is a relatively slow heat transport mechanism compared to accretion, it is only effective on timescales of the order of $\sim 10^8$ yr.

As the radius of the growing moon ($R_s$) increases, impactors accelerate ($v_{imp} \propto R_s^2$), leading to more energetic collisions. Consequently, the heat generated by these impacts intensifies towards the end of the accretion phase, potentially melting the ice in several outer layers, as shown by the high temperature rises in the right panel of Fig. \ref{figure:Callisto_example}. Although impacts become more energetic during this stage, the peak temperature occurs primarily in the subsurface rather than at the surface. This is because radiative cooling, which dominates heat transport, prevents surface temperatures from rising significantly. When the temperature increases by $\delta T$, the surface rapidly cools through radiation within a few hundred days. As a result, large impactors are unable to efficiently heat the surface, and the surface temperature is instead determined by the balance of fluxes (see Eq. \ref{equation:Tsurf}).

During accretion, localized melting can lead to the formation of transient melt pools in which rock particles separate from ice and settle to the base, creating a dense, rock-rich layer above a less dense, undifferentiated region (see Section \ref{subsection:Melting&Diff}). As illustrated in Figure \ref{figure:Callisto_example}, the ice–rock mantles of both satellites reach temperatures above $\sim$180 K, resulting in sufficiently low viscosities to enable gravitational instability. Consequently, the dense layer becomes prone to convective overturn on timescales comparable to that of satellite accretion. This overturn contributes both to the satellite’s thermal evolution and to its partial differentiation. In this study, global melting and subsequent differentiation are assumed to occur once the water liquidus is exceeded. Since our simulations follow a Monte Carlo approach, the timing and frequency of impacts are randomly distributed, leading to slight run--to--run variations in the thermal evolution, even under identical input parameters. Specifically, the volume of material exceeding the liquidus may vary by up to 5\% between simulations. To account for this variability, we incorporate a margin of uncertainty into our differentiation criterion, and define global melting as occurring when a small fraction—up to 5\% by volume—of the satellite’s interior surpasses the liquidus, consistent with the threshold adopted by \citet{MONTEUX2014}. Under this criterion, the simulation shown in Figure \ref{figure:Callisto_example}, which yields only 2.2\% melting by volume, suggests that Callisto, unlike Ganymede, likely avoided global melting during accretion for the considered set of parameters.

During the accretion phase, tidal forces can be neglected compared to radiogenic and accretional heating. Using Eq. \ref{equation:tidal}, we find that tidal dissipation is several orders of magnitude lower than radiogenic heating, with radiogenic heating $Q_{\text{rad}} \approx $ 8.0 $\times$ $10^{-6}$ W/m$^3$ at time $t_{\text{start}} = 3$ Myr after CAIs. In contrast, tidal dissipation $Q_{\text{tid}} \approx$ 9.3 $\times$ $10^{-10}$ and 3.9 $\times$ $10^{-9}$ W/m$^3$ for Callisto and Ganymede, respectively. However, during the post-accretion phase, which is beyond the scope of this study, the situation changes. Long-lived radionuclides like uranium and thorium provide heating sources that are comparable to tidal heating, making the latter a more significant contributor to the interior's warming. This additional tidal heating would complement both the heat retained from the accretion phase and the ongoing radiogenic heating.

\label{subsection:parameter_space_exploration}

\begin{figure*}[!ht]
\resizebox{\hsize}{!}{\includegraphics[angle=0,width=5cm]{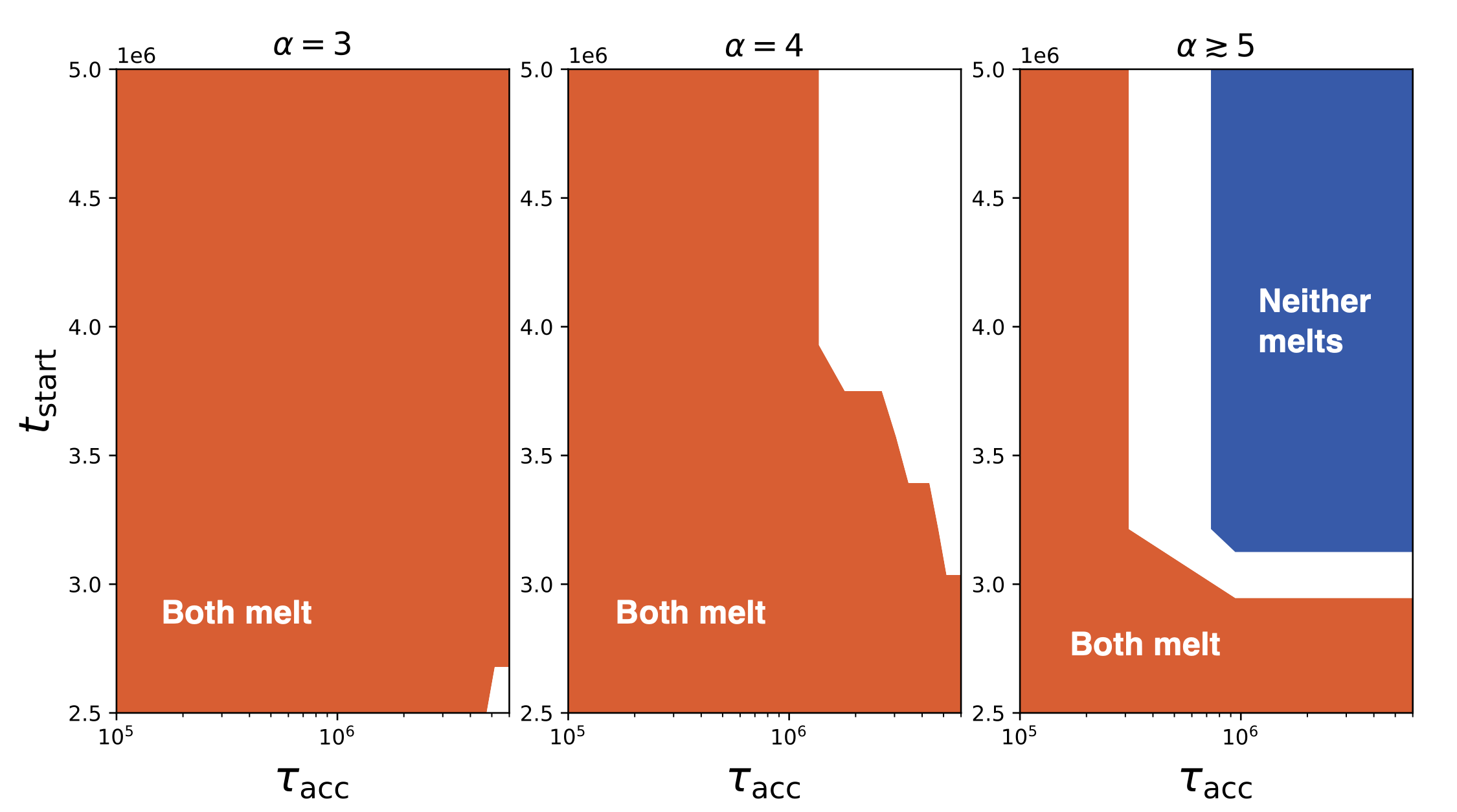}}
\caption{Final states of Callisto and Ganymede as functions of $t_{\text{start}}$, $\tau_{\text{acc}}$, and $\alpha$. The white region indicates where Ganymede undergoes melting, while Callisto remains undifferentiated. From left to right, the panels display increasing values of $\alpha$, ranging from 3 to 5, with a total of 100 simulations per panel. Values equal to 6 are not shown, as the mass fraction distribution remains unchanged between $\alpha = 5$ and $\alpha = 6$, with $\sim 100$ wt\% for small impactors ($x^{si}_{\rm m})$ in both cases. To prevent melting during formation, Callisto must accrete a minimum amount of large impactors ($\alpha \gtrsim 4$), namely less than 30\% of its final mass. For \(\alpha = 4\) (or $\alpha \gtrsim 5$) Callisto’s accretion likely occurred slowly, with $\tau_{\text{acc}} \gtrsim 2$ Myr (or $\tau_{\text{acc}} \gtrsim 0.6$ Myr) and started no earlier than $t_{\text{start}}=3.5$ (or $t_{\text{start}}=3$ Myr) after CAIs. Using the same parameters that prevented Callisto from melting, Ganymede could have differentiated early if \(\alpha = 4\), suggesting that the observed dichotomy may be primordial.}
\label{figure:Param_exploration}
\end{figure*}

Figure \ref{figure:Param_exploration} illustrates the melting behavior of satellites as influenced by the key parameters $\alpha$, $\tau_{\text{acc}}$ and $t_{\text{start}}$. The top two panels, corresponding to a slope $\alpha =$ 3, show that both Ganymede and Callisto undergo global ice melting when large impactors predominantly dominate the size distribution, regardless of the accretion onset time or timescale. To prevent Callisto from melting during its accretion, the population of impactors must be skewed towards smaller bodies ($\alpha \gtrsim$ 4). More precisely, the satellite must accumulate only a small fraction of large impactors, with less than $x^{li}_{\rm m}\sim30$ \% of its final mass. Otherwise, substantial energy will be deposited deep within the moon due to shock heating, and neither prolonged accretion times nor delayed formation would be sufficient to inhibit melting.

\begin{table*}[!tb]
\centering
\caption{Summary of scenarios leading to a primordial dichotomy} 
\begin{tabular}{ll|l}
\hline
\hline
Accretional parameters & Callisto undifferentiated & Ganymede melts \\ 
($\tau_{\text{acc}}$, $t_{\text{start}}$) & & \\
\hline 
Same values     
& \multicolumn{2}{c}{$\alpha \sim 4$, $\tau_{\text{acc}} >$ 2 Myr, $t_{\text{start}} \gtrsim 3.5$ Myr}      \\ 
& \multicolumn{2}{c}{$\alpha \ge 5$, $\tau_{\text{acc}} \in [0.6, 1]$ Myr, $t_{\text{start}} \gtrsim 3$ Myr}      \\ 
\hline
Different values  
& $\alpha=4$: $\tau_{\text{acc}} >$ 2 Myr, $t_{\text{start}} \gtrsim 3.5$ Myr 
& $\alpha=4$: No conditions on $\tau_{\text{acc}}$ and $t_{\text{start}}$ \\ 
& $\alpha\ge5$: $\tau_{\text{acc}} >$ 0.6 Myr, $t_{\text{start}} \gtrsim 3$ Myr 
& $\alpha \ge 5$: $\tau_{\text{acc}} <$ 1 Myr \text{or} $t_{\text{start}} \lesssim 3$ Myr \\
\hline
\end{tabular}
\label{tab:Dichotomy}
\end{table*}

Our calculations suggest that Callisto’s accretion likely occurred slowly ($\tau_{\text{acc}} \gtrsim 2$Myr) and concluded no earlier than 5.5 Myr for $\alpha = 4$. Unlike accretional heating, which effectively melts the ice in the outer layers of the satellites, radiogenic heating primarily affects their deep interiors, as the heating rate depends on the formation time of each layer. When considering both accretional and radiogenic heating, the thermal perturbations caused by impact heating add to the heat from radioisotopes. Consequently, if small particles dominate the impactor population, the period during which accretion occurs without melting could be shorter and begin earlier. Typically,  for $\alpha \gtrsim 5$, we find that Callisto could have formed without melting over a timescale of $\tau_{\text{acc}} \gtrsim 0.6$ Myr, and with $t_{\text{start}} \gtrsim 3$ Myr.

It is also noteworthy that an accretion onset time of less than $t_{\text{start}}=3$ Myr results in global differentiation, regardless of the particle size distribution or accretion timescale. The substantial abundance of $^{26}$Al causes a significant rise in interior temperature, approaching the melting point. As the viscosity of the ice mantle decreases considerably, rock particles may begin to sink towards the center of the protosatellite, creating conditions conducive to slow differentiation. 

The main differences between the formation conditions of Ganymede and Callisto are the disk temperature and the final radius, which result in higher impact velocities for Ganymede ($v_{imp} \propto R_s^2$). Ganymede can only be preserved from melting if the power law slope $\alpha \gtrsim 5$, indicating that only small bodies are accreted. Furthermore, longer accretion timescales or later accretion could not prevent melting if Ganymede is formed from objects larger than a few hundred meters.

Figure \ref{figure:Param_exploration} further explores the parameter space ($\alpha$, $\tau_{\text{acc}}$, $t_{\text{start}}$) that could produce the primordial dichotomy observed in the classic gas-starved disk model \citep{2002CanupWard}, where Ganymede and Callisto are thought to have formed around the same time. The middle panels show that $\alpha =4$ is the most suitable slope to explain this dichotomy. With the same parameters $\tau_{\text{acc}}$ and $t_{\text{start}}$ that prevent Callisto from melting, we find that Ganymede could achieve early differentiation if both satellites accreted a significant number of medium and large-sized bodies (\(\alpha \sim 4\)), over a timescale $\tau_{\text{acc}} \gtrsim 2$ Myr, and before $t_{\text{start}} \gtrsim 3.5$ Myr after CAIs.

\section{\textbf{Discussion}} \label{sec:discussion}


 We have shown that Callisto can remain undifferentiated, despite forming through the accretion of a substantial influx of kilometer-sized impactors --up to 30\% of its final mass-- while simultaneously allowing the complete differentiation of Ganymede. This dichotomy between the two moons naturally emerges during their accretion phases, as both form under identical conditions governed by the same set of parameters ($\tau_{\text{acc}}$, $t_{\text{start}}$, $\alpha$). The main differences in their formation conditions are the disk temperature and the final radius, which result in higher impact velocities for Ganymede. These higher velocities favor more efficient heating and differentiation in Ganymede compared to Callisto. Additional simulations show that the dichotomy between the two moons can also be explained if they accrete with the same mass flux, $\dot{M}_{\rm Sat}$, meaning that Ganymede completes its accretion slightly later than Callisto.

For this dichotomy to be primordial, our results suggest that both moons likely accreted gradually ($\tau_{\text{acc}} > 2$ Myr), starting after $t_{\text{start}} \gtrsim 3.5$ Myr, with $\alpha \sim 4$. The preferred size distribution in this study lies between the two extremes, $\alpha = 1$ (where the impactor mass is concentrated in large bodies) and $\alpha = 6$ (where the impactors are smaller). Consequently, we propose that satellite bodies with radii $r_{\text{imp}} \gtrsim 1$ km were likely present in the Jovian CPD during satellite formation, contributing up to 30\% of the total mass influx. These satellitesimals may represent collisional debris or trapped planetesimals that have lost a significant fraction of their mass through ablation \citep{Ronnet_Johansen}, while the bulk of the mass is contained in smaller particles ranging from centimeters to tens of meters.

The values proposed here cannot constrain the size distribution of particles within the PSN at Jupiter’s orbit, as the majority of planetesimals would have been ablated by the gas giant, contributing to a significant reservoir of pebbles within the Jovian subnebula. In this context, the size distribution of the PSN at Jupiter’s orbit was likely modified and evolved towards steeper slopes during CPD accretion, where the mass influx was dominated by small particles. If the building blocks of satellites formed during the earliest stages of the CPD, our preferred value of $\alpha \sim 4$ aligns with the pebble accretion scenario \citep{Ronnet_2017,Ronnet_Johansen}, challenging the view that collisions between satellitesimals dominate the growth process.

{\citet{Batygin_2020} did not include pebble accretion in their model, focusing instead on oligarchic growth, which resulted in a size-frequency distribution with a significantly flatter slope than $\alpha = 3$. Even when considering debris from disk collisions, their model predicts an almost complete mass fraction of large impactors, leading to global melting of both moons. However, the size distribution inferred in this study may represent a second-generation distribution, in which the observed satellites accreted later from an evolved size-frequency distribution in which particles underwent both effective growth and collisional erosion. This interpretation is consistent with the measured value for main belt asteroids, $\alpha \sim 3.5$ \citep{Zellner1979}, and suggests that oligarchic growth cannot be ruled out for the initial formation of large structures that were subsequently processed and eroded by collisional debris.


Ganymede is unlikely to avoid melting during accretion. Due to its larger radius and higher disk temperature compared to Callisto, resulting from its closer proximity to Jupiter, Ganymede can only prevent melting by accreting slowly ($\tau_{\text{acc}} \gtrsim 1$ Myr) and from small particles with radii $r_{\text{imp}}$ around 100 meters or less (see Table \ref{tab:Dichotomy}). Accreting even a small number of large impactors buries energy deep within the moon, where it cannot efficiently dissipate to the surface. Consequently, neither prolonged accretion times nor delayed formation would be sufficient to prevent melting, as removing the energy deposited at these depths would require a timescale of about $10^8$ years. This finding contrasts with the traditional view that larger impactors require longer accretion times or delayed formation to avoid melting, as discussed in \cite{2008Barr} and \cite{Batygin_2020}.

One could envision a scenario where Ganymede and Callisto formed exclusively from small bodies with radii of $r_{\text{imp}} \lesssim 100$ meters and had an accretion timescale of $\tau_{\text{acc}} \gtrsim 1$ Myr, which aligns well with the Canup \& Ward model \citep{2002CanupWard, 2008Barr}. However, the possibility of Ganymede and Callisto accreting only small particles is low, given the presence of km-sized satellitesimals in the CPD. \cite{Ronnet_Johansen} demonstrated that about 10\% of planetesimals captured from the PSN avoided complete mass loss during ablation, retaining significant sizes of several kilometers. However, the current size-frequency distribution of craters on satellite surfaces may not provide additional constraints on impactor sizes during accretion, as no major satellites have surface ages that extend back to the circumplanetary disk era \citep{Bottke_2024}. The impact record was likely erased due to resurfacing events and subsequent cratering by heliocentric impactors, with Jupiter-family comets accounting for over 90\% of the craters observed on the Galilean satellites \citep{ZAHNLE1998202}.

Nevertheless, if both Callisto and Ganymede accreted small bodies, the dichotomy between the two moons may eventually result from divergent evolutionary pathways. \cite{2010Barr} proposed that the late heavy bombardment (LHB) could explain the differences, with Ganymede possibly experiencing more energetic impacts than Callisto due to probabilistic variations. However, a recent study by \cite{Bottke_2024} questions this scenario, as their estimated mass flux of impactors during the LHB is five times lower than that assumed by \cite{2010Barr}, reducing the likelihood of Ganymede melting due to impacts. Similarly, \cite{NIMMO2012} explain that the mass delivered to the satellites during the post-accretional bombardment is too high to be consistent with observations. To reconcile this discrepancy, the mass inflow from the destabilized population must be reduced by a factor of 10. Otherwise, Mimas, Miranda, and Enceladus would lose all of their volatiles, assuming these icy bodies formed well before the bombardment.

Resonances could also explain the surface differences between Ganymede and Callisto. \cite{Showman1997} and \cite{Peale2002} proposed the existence of an evolutionary path to the Laplace resonance, where Ganymede's eccentricity increases to 0.07 before settling to its present value. Consequently, \cite{Showman1997_2} suggested that the significant tidal heating Ganymede experienced during this period would have led to resurfacing and the formation of the young, bright terrains observed today. However, the authors assume that Ganymede had a fully differentiated initial structure with a warm icy mantle before entering the Laplace resonance. This result may be consistent with our study, as the paper also suggests that differentiation may have occurred early, during the accretion phase, and before the formation of the dark terrains.

Given the uncertainty in the gas density distribution within the CPD, which directly influences the formation timescales, Callisto and Ganymede could also have fundamentally different accretion parameters, especially $\tau_{\text{acc}}$ and $t_{\text{start}}$. In this situation, the dichotomy could still be primordial, as illustrated by the various scenarios summarized in Table \ref{tab:Dichotomy}.
In addition, Europa and Io probably accreted in the same satellitesimal feeding region as Ganymede \citep{Batygin_2020}, implying similar or even shorter accretion timescales ($\tau_{\text{acc}} \sim 10^3$--$10^4$ yr) \citep{estrada2009formation}. In this scenario, melting of the two inner Galilean moons seems inevitable, probably leading to rock-iron differentiation \citep{Trinh_etal_2023}.



The onset time of Callisto's accretion, $t_{\text{start}}$, can also serve as a constraint on the timing of Jupiter's formation. In the scenarios outlined in Table \ref{tab:Dichotomy}, we find that $t_{\text{start}}$ for Callisto exceeds 3.5 Myr for $\alpha=5$ and 3 Myr for $\alpha=4$. Assuming that the formation of Jupiter ended with the formation of the Jovian CPD and that Callisto began accreting immediately thereafter, the formation of Jupiter probably ended no earlier than $\sim 3$ Myr and 3.5 Myr after CAIs for $\alpha=4$ and $\alpha=5$, respectively. This result supports the late formation hypothesis for Jupiter, suggesting that the gas giant continued to accrete material from the PSN until about 3--5 Myr after the formation of the Solar System \citep{Kruijer2017,2020Kleine}. 

Here, our temperature profiles are based on the disk model described in \cite{Schneeberger2025}. The objective of this study was to couple accretional heating with the disk temperature evolution $T_e(t)$, to examine how the cooling of the disk at the end of the  accretion phase impacts satellite formation, rather than introducing new free parameters or assuming constant temperature values throughout the satellite formation process. However, disk temperatures can exhibit significant variation across different models, with key influences from its opacity and the rate of material inflow. While these temperature variations could potentially impact the accretional temperature profiles, we find that this effect becomes less pronounced when the number of small particles decreases ($\alpha < 4$), as the surface would experience less heating compared to the interior, making it less sensitive to ambient temperature changes. Thus, if a high number of large impactors are present in the Jovian CPD, we expect that the temperature profiles inside protosatellites may be similar to the results presented in this work.

Furthermore, our model assumes in situ formation of both moons, although inward migration of the satellites may have occurred during their formation. According to \cite{2006CanupWard}, the satellites we observe today represent the last generation that likely avoided Type I migration, since they formed after the material inflow ceased. Specifically, \cite{2006CanupWard}'s simulations suggest that Callisto analogues migrated inward by less than 5 Jupiter radii, an effect that would have had only a minor impact on the disk temperature profiles. In contrast, \cite{MOSQUEIRA_a} predicts that Callisto's building blocks originated in the cold outer regions of the disk, about 100 Jupiter radii. However, the satellite embedded model of \cite{MOSQUEIRA_a, MOSQUEIRA_b} predicts a significant fraction of impactors with radii in the 1--100 km range. As noted earlier, within this typical size range, the accretion temperature profiles are not significantly affected by the disk temperature, since most of the energy is retained in the interior of the satellite\footnote{The assumption of an impactor size distribution containing only a few 100 km-sized satellitesimals would be also sufficient to induce significant differentiation of Callisto, in contradiction with the gravity data.}. Therefore, even if Type I migration had occurred during satellite formation at a lower disk temperature, the results should not be significantly different from those presented here.

This study does not explicitly model water transport or water–rock separation, as we assume that differentiation is initiated once a few percent of the material exceeds the melting point of water ice. As discussed in Section \ref{subsection:Melting&Diff}, impacts can produce localized melt pools that rapidly refreeze over timescales of several tens of thousands of years, depending on the melt volume. Given that melting occurs near the end of accretion and is predominantly confined to the surface, we do not expect substantial melt migration before refreezing is complete.

Our model does not include plastic work heating - the conversion of mechanical energy to heat due to irreversible plastic deformation, where a fraction of the plastic strain energy is dissipated as heat. Recent studies have shown that plastic deformation during decompression can contribute to energy deposition in impacts, especially in low-velocity collisions \citep{Kurosawa2018, Kurosawa2021}. This mechanism could lead to localized surface melting and devolatilization, potentially influencing the thermal evolution of icy bodies such as Ganymede and Callisto. However, incorporating plastic work heating into our model would introduce additional complexity by requiring the tracking of local deformation, which falls outside the scope of this study. Instead, we prioritize shock wave heating due to its greater impact on thermal changes during collisions. Future work that integrates plastic work heating into accretion models could provide a more accurate representation of impact processes and provide deeper insights into the complex thermal dynamics of icy-rock environments.

\begingroup
\renewcommand\linenumberfont[1]{}
\begin{acknowledgments}
OM acknowledges support from CNES. This research holds as part of the project FACOM (ANR-22-CE49-0005-01 ACT) and has benefited from a funding provided by l’Agence Nationale de la Recherche (ANR) under the Generic Call for Proposals 2022.
VH acknowledges support from the French government under the France 2030 investment plan, as part of the Initiative d’Excellence d’Aix-Marseille Université – A*MIDEX AMX-22-CPJ-04.
\end{acknowledgments}
\endgroup

\bibliography{sample631}{}

\begin{thebibliography}{}
\expandafter\ifx\csname natexlab\endcsname\relax\def\natexlab#1{#1}\fi
\providecommand{\url}[1]{\href{#1}{#1}}
\providecommand{\dodoi}[1]{doi:~\href{http://doi.org/#1}{\nolinkurl{#1}}}
\providecommand{\doeprint}[1]{\href{http://ascl.net/#1}{\nolinkurl{http://ascl.net/#1}}}
\providecommand{\doarXiv}[1]{\href{https://arxiv.org/abs/#1}{\nolinkurl{https://arxiv.org/abs/#1}}}

\bibitem[{Anderson {et~al.}(2001)Anderson, Jacobson, McElrath, Moore, Schubert, \& Thomas}]{ANDERSON2001157}
Anderson, J., Jacobson, R., McElrath, T., {et~al.} 2001, Icarus, 153, 157, \dodoi{https://doi.org/10.1006/icar.2001.6664}

\bibitem[{{Barr} \& {Canup}(2008)}]{2008Barr}
{Barr}, A., \& {Canup}, R. 2008, Icarus, 198, 163

\bibitem[{{Barr} \& {Canup}(2010{\natexlab{a}})}]{2010Barr}
---. 2010{\natexlab{a}}, Icarus, 209, 858

\bibitem[{{Barr} \& {Canup}(2010{\natexlab{b}})}]{2010Barr_Dichotomy}
{Barr}, A.~C., \& {Canup}, R.~M. 2010{\natexlab{b}}, Nature Geoscience, 3, 164, \dodoi{10.1038/ngeo746}

\bibitem[{Barr {et~al.}(2010)Barr, Citron, \& Canup}]{BARR2010_Titan}
Barr, A.~C., Citron, R.~I., \& Canup, R.~M. 2010, Icarus, 209, 858, \dodoi{https://doi.org/10.1016/j.icarus.2010.05.028}

\bibitem[{Batygin \& Morbidelli(2020)}]{Batygin_2020}
Batygin, K., \& Morbidelli, A. 2020, The Astrophysical Journal, 894, 143, \dodoi{10.3847/1538-4357/ab8937}

\bibitem[{Bottke {et~al.}(2024)Bottke, Vokrouhlick, Nesvorn, Marschall, Morbidelli, Deienno, Marchi, Kirchoff, Dones, \& Levison}]{Bottke_2024}
Bottke, W.~F., Vokrouhlick, D., Nesvorn, D., {et~al.} 2024, The Planetary Science Journal, 5, 88, \dodoi{10.3847/PSJ/ad29f4}

\bibitem[{{Canup} \& {Ward}(2002)}]{2002CanupWard}
{Canup}, R., \& {Ward}, W. 2002, ApJ, 124, 3404, \dodoi{https://iopscience.iop.org/article/10.1086/344684/pdf}

\bibitem[{{Canup} \& {Ward}(2006)}]{2006CanupWard}
{Canup}, R.~M., \& {Ward}, W.~R. 2006, \nat, 441, 834, \dodoi{10.1038/nature04860}

\bibitem[{{Carslaw} \& {Jaeger}(1959)}]{Carslaw1959}
{Carslaw}, H.~S., \& {Jaeger}, J.~C. 1959, {Conduction of heat in solids}

\bibitem[{Castillo-Rogez {et~al.}(2007)Castillo-Rogez, Matson, Sotin, Johnson, Lunine, \& Thomas}]{CASTILLOROGEZ2007179}
Castillo-Rogez, J., Matson, D., Sotin, C., {et~al.} 2007, Icarus, 190, 179, \dodoi{https://doi.org/10.1016/j.icarus.2007.02.018}

\bibitem[{Chivers {et~al.}(2021)Chivers, Buffo, \& Schmidt}]{Chivers_etal_2021}
Chivers, C.~J., Buffo, J.~J., \& Schmidt, B.~E. 2021, Journal of Geophysical Research: Planets, 126, e2020JE006692, \dodoi{https://doi.org/10.1029/2020JE006692}

\bibitem[{Croft(1982)}]{Croft1982}
Croft, S.~K. 1982, in {Geological Implications of Impacts of Large Asteroids and Comets on the Earth} (Geological Society of America), \dodoi{10.1130/SPE190-p143}

\bibitem[{{Downey} {et~al.}(2020){Downey}, {Nimmo}, \& {Matsuyama}}]{2020Downey}
{Downey}, B., {Nimmo}, F., \& {Matsuyama}, I. 2020, MNRAS, 499, 40

\bibitem[{Estrada {et~al.}(2009)Estrada, Mosqueira, Lissauer, D'Angelo, \& Cruikshank}]{estrada2009formation}
Estrada, P.~R., Mosqueira, I., Lissauer, J.~J., D'Angelo, G., \& Cruikshank, D.~P. 2009, Formation of Jupiter and Conditions for Accretion of the Galilean Satellites.
\newblock \doarXiv{0809.1418}

\bibitem[{Heller {et~al.}(2015)Heller, Marleau, \& Pudritz}]{heller2015}
Heller, R., Marleau, G.~D., \& Pudritz, R.~E. 2015, 579, L4, \dodoi{10.1051/0004-6361/201526348}

\bibitem[{Hesse \& Castillo-Rogez(2019)}]{Hesse&Castillo2019}
Hesse, M.~A., \& Castillo-Rogez, J.~C. 2019, Geophysical Research Letters, 46, 1213, \dodoi{https://doi.org/10.1029/2018GL080327}

\bibitem[{Kivelson {et~al.}(1998)Kivelson, Warnecke, Bennett, Joy, Khurana, Linker, Russell, Walker, \& Polanskey}]{Kivelson1988}
Kivelson, M.~G., Warnecke, J., Bennett, L., {et~al.} 1998, Journal of Geophysical Research: Planets, 103, 19963, \dodoi{https://doi.org/10.1029/98JE00227}

\bibitem[{{Kleine} {et~al.}(2020){Kleine}, {Budde}, {Burkhardt}, {Kruijer}, {Worsham}, {Morbidelli}, \& {Nimmo}}]{2020Kleine}
{Kleine}, T., {Budde}, G., {Burkhardt}, C., {et~al.} 2020, ApJ, 216, 3404, \dodoi{https://doi.org/10.1007/s11214-020-00675-w}

\bibitem[{Klinger(1980)}]{Klinger1980}
Klinger, J. 1980, Science, 209, 271, \dodoi{10.1126/science.209.4453.271}

\bibitem[{Kruijer {et~al.}(2017)Kruijer, Burkhardt, Budde, \& Kleine}]{Kruijer2017}
Kruijer, T.~S., Burkhardt, C., Budde, G., \& Kleine, T. 2017, Proceedings of the National Academy of Sciences, 114, 6712, \dodoi{10.1073/pnas.1704461114}

\bibitem[{Kurosawa \& Genda(2018)}]{Kurosawa2018}
Kurosawa, K., \& Genda, H. 2018, Geophysical Research Letters, 45, 620, \dodoi{https://doi.org/10.1002/2017GL076285}

\bibitem[{Kurosawa {et~al.}(2021)Kurosawa, Genda, Azuma, \& Okazaki}]{Kurosawa2021}
Kurosawa, K., Genda, H., Azuma, S., \& Okazaki, K. 2021, Geophysical Research Letters, 48, e2020GL091130, \dodoi{https://doi.org/10.1029/2020GL091130}

\bibitem[{Leliwa-Kopysty\'nski {et~al.}(2002)Leliwa-Kopysty\'nski, Maruyama, \& Nakajima}]{Leliwa2002}
Leliwa-Kopysty\'nski, J., Maruyama, M., \& Nakajima, T. 2002, Icarus, 159, 518, \dodoi{https://doi.org/10.1006/icar.2002.6932}

\bibitem[{Lewis(1971)}]{LEWIS1971}
Lewis, J.~S. 1971, Icarus, 15, 174, \dodoi{https://doi.org/10.1016/0019-1035(71)90072-8}

\bibitem[{Lodders(2003)}]{Lodders_2003}
Lodders, K. 2003, The Astrophysical Journal, 591, 1220, \dodoi{10.1086/375492}

\bibitem[{Lodders(2019)}]{lodders2019}
---. 2019, Solar {{Elemental Abundances}}, \dodoi{10.48550/arXiv.1912.00844}

\bibitem[{Makalkin \& Dorofeeva(1995)}]{makalkin1995}
Makalkin, A.~B., \& Dorofeeva, V.~A. 1995, Astronomicheskii Vestnik, 29, 99

\bibitem[{Makalkin \& Dorofeeva(2014)}]{makalkin2014}
---. 2014, Solar System Research, 48, 62, \dodoi{10.1134/S0038094614010067}

\bibitem[{Malhotra(1991)}]{Malhotra1991}
Malhotra, R. 1991, Icarus, 94, 399, \dodoi{10.1016/0019-1035(91)90237-N}

\bibitem[{{McCord} \& {Sotin}(2005)}]{McCord&Sotin2005}
{McCord}, T.~B., \& {Sotin}, C. 2005, Journal of Geophysical Research (Planets), 110, E05009, \dodoi{10.1029/2004JE002244}

\bibitem[{McKinnon(1997)}]{MCKINNON1997540}
McKinnon, W.~B. 1997, Icarus, 130, 540, \dodoi{https://doi.org/10.1006/icar.1997.5826}

\bibitem[{Michaut \& Manga(2014)}]{Michaut}
Michaut, C., \& Manga, M. 2014, Journal of Geophysical Research: Planets, 119, 550, \dodoi{https://doi.org/10.1002/2013JE004558}

\bibitem[{Mishra \& Chaussidon(2014)}]{MISHRA201490}
Mishra, R.~K., \& Chaussidon, M. 2014, Earth and Planetary Science Letters, 398, 90, \dodoi{https://doi.org/10.1016/j.epsl.2014.04.032}

\bibitem[{Monnereau {et~al.}(2013)Monnereau, Toplis, Baratoux, \& Guignard}]{MONNEREAU2013}
Monnereau, M., Toplis, M.~J., Baratoux, D., \& Guignard, J. 2013, Geochimica et Cosmochimica Acta, 119, 302, \dodoi{https://doi.org/10.1016/j.gca.2013.05.035}

\bibitem[{Monteux {et~al.}(2007)Monteux, Coltice, Dubuffet, \& Ricard}]{Monteux2007}
Monteux, J., Coltice, N., Dubuffet, F., \& Ricard, Y. 2007, Geophysical Research Letters, 34, \dodoi{https://doi.org/10.1029/2007GL031635}

\bibitem[{Monteux {et~al.}(2014)Monteux, Tobie, Choblet, \& {Le Feuvre}}]{MONTEUX2014}
Monteux, J., Tobie, G., Choblet, G., \& {Le Feuvre}, M. 2014, Icarus, 237, 377, \dodoi{https://doi.org/10.1016/j.icarus.2014.04.041}

\bibitem[{Morris(2001)}]{Petrenko1999}
Morris, E.~M. 2001, Antarctic Science, 13, 350–351, \dodoi{10.1017/S0954102001220483}

\bibitem[{Mosqueira \& Estrada(2003{\natexlab{a}})}]{MOSQUEIRA_a}
Mosqueira, I., \& Estrada, P.~R. 2003{\natexlab{a}}, Icarus, 163, 198, \dodoi{https://doi.org/10.1016/S0019-1035(03)00076-9}

\bibitem[{Mosqueira \& Estrada(2003{\natexlab{b}})}]{MOSQUEIRA_b}
---. 2003{\natexlab{b}}, Icarus, 163, 232, \dodoi{https://doi.org/10.1016/S0019-1035(03)00077-0}

\bibitem[{Mousis {et~al.}(2002)Mousis, Pargamin, Grasset, \& Sotin}]{Mousis2002}
Mousis, O., Pargamin, J., Grasset, O., \& Sotin, C. 2002, Geophysical Research Letters, 29, 45, \dodoi{https://doi.org/10.1029/2002GL015812}

\bibitem[{Nagel {et~al.}(2004)Nagel, Breuer, \& Spohn}]{NAGEL2004402}
Nagel, K., Breuer, D., \& Spohn, T. 2004, Icarus, 169, 402, \dodoi{https://doi.org/10.1016/j.icarus.2003.12.019}

\bibitem[{Naseem {et~al.}(2023)Naseem, Neveu, Howell, Lesage, Melwani~Daswani, \& Vance}]{Naseem_2023}
Naseem, M., Neveu, M., Howell, S., {et~al.} 2023, The Planetary Science Journal, 4, 181, \dodoi{10.3847/PSJ/ace5a2}

\bibitem[{Nimmo \& Korycansky(2012)}]{NIMMO2012}
Nimmo, F., \& Korycansky, D. 2012, Icarus, 219, 508, \dodoi{https://doi.org/10.1016/j.icarus.2012.01.016}

\bibitem[{{Okeefe} \& {Ahrens}(1977)}]{Okeefe1977}
{Okeefe}, J.~D., \& {Ahrens}, T.~J. 1977, Lunar and Planetary Science Conference Proceedings, 3, 3357

\bibitem[{Peale \& Lee(2002)}]{Peale2002}
Peale, S.~J., \& Lee, M.~H. 2002, Science, 298, 593, \dodoi{10.1126/science.1076557}

\bibitem[{{Ronnet} \& {Johansen}(2020)}]{Ronnet_Johansen}
{Ronnet}, T., \& {Johansen}, A. 2020, \aap, 633, A93, \dodoi{10.1051/0004-6361/201936804}

\bibitem[{Ronnet {et~al.}(2017)Ronnet, Mousis, \& Vernazza}]{Ronnet_2017}
Ronnet, T., Mousis, O., \& Vernazza, P. 2017, The Astrophysical Journal, 845, 92, \dodoi{10.3847/1538-4357/aa80e6}

\bibitem[{Rubin {et~al.}(2014)Rubin, Desch, \& Neveu}]{RUBIN2014}
Rubin, M.~E., Desch, S.~J., \& Neveu, M. 2014, Icarus, 236, 122, \dodoi{https://doi.org/10.1016/j.icarus.2014.03.047}

\bibitem[{Schneeberger \& Mousis(2025)}]{Schneeberger2025}
Schneeberger, A., \& Mousis, O. 2025, The Planetary Science Journal, 6, 23, \dodoi{10.3847/PSJ/ad9de1}

\bibitem[{Schubert {et~al.}(1981)Schubert, Stevenson, \& Ellsworth}]{SCHUBERT198146}
Schubert, G., Stevenson, D., \& Ellsworth, K. 1981, Icarus, 47, 46, \dodoi{https://doi.org/10.1016/0019-1035(81)90090-7}

\bibitem[{{Senshu} {et~al.}(2002){Senshu}, {Kuramoto}, \& {Matsui}}]{Senshu2002}
{Senshu}, H., {Kuramoto}, K., \& {Matsui}, T. 2002, Journal of Geophysical Research (Planets), 107, 5118, \dodoi{10.1029/2001JE001819}

\bibitem[{Shoji \& Kurita(2014)}]{Shoji&Kurita2014}
Shoji, D., \& Kurita, K. 2014, Journal of Geophysical Research: Planets, 119, 2457, \dodoi{https://doi.org/10.1002/2014JE004695}

\bibitem[{{Showman} \& {Malhotra}(1997)}]{Showman1997}
{Showman}, A.~P., \& {Malhotra}, R. 1997, \icarus, 127, 93, \dodoi{10.1006/icar.1996.5669}

\bibitem[{{Showman} {et~al.}(1997){Showman}, {Stevenson}, \& {Malhotra}}]{Showman1997_2}
{Showman}, A.~P., {Stevenson}, D.~J., \& {Malhotra}, R. 1997, \icarus, 129, 367, \dodoi{10.1006/icar.1997.5778}

\bibitem[{{Spohn} \& {Schubert}(2003)}]{Spohn_Schubert_2003}
{Spohn}, T., \& {Schubert}, G. 2003, \icarus, 161, 456, \dodoi{10.1016/S0019-1035(02)00048-9}

\bibitem[{Squyres {et~al.}(1988)Squyres, Reynolds, Summers, \& Shung}]{Squyres}
Squyres, S.~W., Reynolds, R.~T., Summers, A.~L., \& Shung, F. 1988, Journal of Geophysical Research: Solid Earth, 93, 8779, \dodoi{https://doi.org/10.1029/JB093iB08p08779}

\bibitem[{Szul{\'a}gyi {et~al.}(2016)Szul{\'a}gyi, Masset, Lega, Crida, Morbidelli, \& Guillot}]{szulagyi2016}
Szul{\'a}gyi, J., Masset, F., Lega, E., {et~al.} 2016, Monthly Notices of the Royal Astronomical Society, 460, 2853, \dodoi{10.1093/mnras/stw1160}

\bibitem[{Thrane {et~al.}(2006)Thrane, Bizzarro, \& Baker}]{Thrane2006}
Thrane, K., Bizzarro, M., \& Baker, J. 2006, Astrophysical Journal, L159

\bibitem[{Tonks {et~al.}(2016)Tonks, Pierazzo, \& Melosh}]{tonks2016impactinduceddifferentiationicybodies}
Tonks, W.~B., Pierazzo, E., \& Melosh, H.~J. 2016, Impact-induced differentiation in icy bodies.
\newblock \doarXiv{1607.01282}

\bibitem[{Trinh {et~al.}(2023)Trinh, Bierson, \& O'Rourke}]{Trinh_etal_2023}
Trinh, K.~T., Bierson, C.~J., \& O'Rourke, J.~G. 2023, Science Advances, 9, eadf3955, \dodoi{10.1126/sciadv.adf3955}

\bibitem[{Waite {et~al.}(2006)Waite, Gilbert, Winters, \& Mason}]{Waite2006}
Waite, W.~F., Gilbert, L.~Y., Winters, W.~J., \& Mason, D.~H. 2006, Review of Scientific Instruments, 77, 044904, \dodoi{10.1063/1.2194481}

\bibitem[{{Wisdom}(2004)}]{Wisdom2004}
{Wisdom}, J. 2004, \aj, 128, 484, \dodoi{10.1086/421360}

\bibitem[{Wisdom(2008)}]{WISDOM2007}
Wisdom, J. 2008, Icarus, 193, 637, \dodoi{https://doi.org/10.1016/j.icarus.2007.09.002}

\bibitem[{Yomogida \& Matsui(1983)}]{Yomogida1983}
Yomogida, K., \& Matsui, T. 1983, Journal of Geophysical Research: Solid Earth, 88, 9513, \dodoi{https://doi.org/10.1029/JB088iB11p09513}

\bibitem[{Zahnle {et~al.}(1998)Zahnle, Dones, \& Levison}]{ZAHNLE1998202}
Zahnle, K., Dones, L., \& Levison, H.~F. 1998, Icarus, 136, 202, \dodoi{https://doi.org/10.1006/icar.1998.6015}

\bibitem[{{Zellner}(1979)}]{Zellner1979}
{Zellner}, B. 1979, in Asteroids, ed. T.~{Gehrels} \& M.~S. {Matthews}, 783--806

\end{thebibliography}
\bibliographystyle{aasjournal}



\end{document}